\documentclass[12pt]{article}
\usepackage{amssymb}
\usepackage{amsmath}

\newtheorem{proposition}{Proposition}

\newtheorem{definition}{Definition}
\begin{document}

\title{Beyond Lyapunov: Ergodic parameters and dynamical complexity}
\author{\textit{R. Vilela Mendes}\thanks{%
vilela@cii.fc.ul.pt, http://label2.ist.utl.pt/vilela/} \\
%EndAName
CMAF\thanks{%
Av. Prof. Gama Pinto 2, 1649-003 Lisboa, Portugal}, Complexo
Interdisciplinar, UL;\\
IPFN - Instituto Superior T\'{e}cnico}
\date{ }
\maketitle

\begin{abstract}
Ergodic parameters like the Lyapunov and the conditional exponents are
global functions of the invariant measure, but the invariant measure itself
contains more information. A more complete characterization of the dynamics
by new families of ergodic parameters is discussed, as well as their
relation to the dynamical R\'{e}nyi entropies and measures of complexity. A
generalization of the Pesin formula is derived which holds under weak
correlation conditions.
\end{abstract}



\section{Introduction}

Ergodic parameters associated to an invariant measure play a central role in
the characterization of dynamical systems. In addition to rigorous notions
of chaos, they also provide indicators of self-organization \cite{Vilela4},
sufficient conditions for self-organized criticality \cite{BakLyap} and a
characterization of topological transitions in networks \cite{SWTa}.

The Lyapunov \cite{Lyapunov} and the conditional exponents \cite{Pecora} 
\cite{Vilela1}, are global functions of the invariant measure. However, the
invariant measure itself contains more information. Ergodic parameters,
being obtained from infinite-time limits, are averages of local fluctuating
quantities. The quantity describing the fluctuations is again an ergodic
parameter and the same reasoning applies in turn to its fluctuations, etc. 
\cite{Ruelle}. Therefore, unless the fluctuations are fluctuations of a
Gaussian random variable, to fully characterize the invariant measure, a
much larger set of parameters is in general needed.

The task of constructing a larger set of ergodic parameters has already been
addressed by several authors. For example, Farmer, Sidorowich and Dressler 
\cite{Sidorowich} \cite{Dressler} proposed to use infinite-time limits of
higher order derivatives of the map. The existence status of these limits is
weaker \cite{Taylor} (convergence in probability) than for the case of
Lyapunov and conditional exponents. Also, whenever they exist, it turns out
that they are simple functions of the Lyapunov exponents and therefore
provide no new information. Other ergodic parameters, independent from the
Lyapunov exponents, were proposed by several authors, either in the form of
generalized entropies \cite{Fujisaka} \cite{Benzi} \cite{Grassberger1}, as
moments of the local fluctuations of the tangent vectors \cite{Crisanti} 
\cite{Vaienti} \cite{Fujisaka} \cite{Froeschle} \cite{Vanneste} \cite%
{Oliveira} or from the eigenvalues of the Hessian in a variational
formulation .

In this paper, a general cocycle formulation is used which allows to
describe in a unified way the generalized ergodic parameters that have been
proposed in the past as well as a new ergodic parameter that subsumes all
the information on the statistics of local fluctuations of the expansion
rate. That multi-point correlations should also be taken into account in the
ergodic descrition of dynamical systems is pointed out.

Then, in Section 3, the moments of the local expansion rate are related to
generalized entropies and in Section 4 one discusses how the ergodic
parameters may be used to characterize and quantify the notions of
complexity and dynamical self-organization.

\section{Generalized ergodic parameters. A cocycle approach}

Let $f:M\rightarrow M$ be a measure-preserving transformation of a Lebesgue
space $\left( M,\mathcal{B},\mu \right) $. For any measurable function $%
g:M\rightarrow GL\left( N,\mathbb{R}\right) $ and $x\in M$ define 
\begin{equation}
C\left( x,n\right) =g\left( f^{n-1}\left( x\right) \right) \cdots g\left(
x\right)  \label{2.1}
\end{equation}%
and $C\left( x,0\right) =$Id . Then 
\begin{equation}
C\left( x,n+k\right) =C\left( f^{k}\left( x\right) ,n\right) C\left(
x,k\right)  \label{2.2}
\end{equation}%
and any measurable function $C:M\times \mathbb{Z}\rightarrow GL\left( N,%
\mathbb{R}\right) $ satisfying (\ref{2.2}) is called a \textit{cocycle}
(over $f$). Any cocycle has the form (\ref{2.1}) and the map $g$ is called
the \textit{generator} of the cocycle.

The Oseledets multiplicative ergodic theorem \cite{Oseledec} is a powerful
result insuring the existence of some infinite-time limits associated to a
cocycle $C$. Namely, if 
\begin{equation}
\ln _{+}\left\Vert g\left( x\right) \right\Vert \in L^{1}\left( M,\mu \right)
\label{2.3}
\end{equation}%
then :

(i) there is a decomposition of $\mathbb{R}^{N}$%
\begin{equation}
\mathbb{R}^{N}=\oplus _{i=1}^{k\left( x\right) }E_{i}\left( x\right)
\label{2.4}
\end{equation}
invariant under $C\left( x,n\right) $,

(ii) and the limits 
\begin{equation}
\lim_{n\rightarrow \infty }\frac{1}{n}\ln \frac{\left\| C\left( x,n\right)
v\right\| }{\left\| v\right\| }=\chi _{i}\left( x\right)  \label{2.5}
\end{equation}
with 
\begin{equation}
\chi _{1}\left( x\right) <\chi _{2}\left( x\right) <\cdots <\chi _{k\left(
x\right) }\left( x\right)  \label{2.6}
\end{equation}
exist uniformly in $v\in E_{i}\left( x\right) \diagdown \{0\}$.

If the generator of the cocyle is 
\begin{equation}
g_{1}\left( x\right) =Df\left( x\right) =\exp \left( \ln \left( Df\left(
x\right) \right) \right)  \label{2.7}
\end{equation}%
the quantities $\chi _{i}\left( x\right) $ are the usual Lyapunov exponents
of the dynamics $f$. If the full Jacobian $Df$ is replaced by partial blocks
of $Df$ one obtains the conditional exponents \cite{Pecora} \cite{Vilela1}.
However, provided that the integrability condition (\ref{2.3}) is satisfied,
Oseledets' theorem applies to any other linear cocycle extension of $f$.

\begin{definition}
The \textit{Lyapunov fluctuation moments} $\chi _{i}^{(p)}\left( x\right) $
are defined as the limits (\ref{2.5}) when the generator of the cocycle is 
\begin{equation}
g_{p}\left( x\right) =\exp \left( \ln _{+}^{p}\left( Df\left( x\right)
\right) \right)  \label{2.8}
\end{equation}
\end{definition}

The definition of the logarithm in (\ref{2.8}) should be understood in the
framework of the Oseledets-Pesin $\varepsilon -$reduction theorem \cite%
{Pesin1} \cite{Katok}. Namely, under the measurability conditions of the
Oseledets theorem, for any $\varepsilon >0$ there is an invertible map $%
\Gamma _{\varepsilon }\left( x\right) :M\rightarrow GL\left( N,\mathbb{R}%
\right) $ such that the generator 
\begin{equation*}
g_{\varepsilon }\left( x\right) =\Gamma ^{-1}\left( f\left( x\right) \right)
g\left( x\right) \Gamma \left( x\right)
\end{equation*}%
has block form, in each block 
\begin{equation*}
e^{\chi _{i}\left( x\right) -\varepsilon }\leq \left\Vert g_{\varepsilon
}^{i}\left( x\right) v\right\Vert \leq e^{\chi _{i}\left( x\right)
+\varepsilon }
\end{equation*}%
and it generates a cocycle $C_{\varepsilon }\left( x,n\right) $ equivalent
to $C\left( x,n\right) $. The $\ln _{+}$ in (\ref{2.8}) is therefore
computed without ambiguity in each block and one sees that the limit%
\begin{equation*}
\chi _{i}^{(p)}\left( x\right) =\lim_{n\rightarrow \infty }\frac{1}{n}\ln 
\frac{\left\Vert g_{p}\left( f^{n-1}\left( x\right) \right) \cdots
g_{p}\left( x\right) v\right\Vert }{\left\Vert v\right\Vert }
\end{equation*}%
is an ergodic average of the $p-$moment of the local (positive) expansion
rate.

As a consequence of the Oseledets multiplicative ergodic theorem, Lyapunov
fluctuation moments $\chi _{i}^{(p)}\left( x\right) $ exist whenever%
\begin{equation}
\ln _{+}\left\Vert g_{p}\left( x\right) \right\Vert \in L^{1}\left( M,\mu
\right)  \label{2.9}
\end{equation}%
This cocycle construction provides a unified description of the fluctuation
ergodic parameters previously considered by several authors \cite{Crisanti} 
\cite{Fujisaka} \cite{Froeschle} \cite{Vanneste} \cite{Oliveira}.

Existence of the limit (\ref{2.9}) depends on the integrability of 
\begin{equation*}
\exp \left( \sum k_{i}\lambda _{i}^{p}\left( x\right) \right)
\end{equation*}%
$\lambda _{i}\left( x\right) $ being the local expansion rate at the point $%
x $ and $k_{i}$ the multiplicity of this particular rate. However, the
expansion rate random variable may fail to have moments for large $p$. In
that case complete characterization of the fluctuations may be obtained by
the ergodic equivalent of the characteristic function.

\begin{definition}
The \textit{Lyapunov characteristic fluctuation function }$C\left( \alpha
\right) $\textit{\ }is defined as the limit (\ref{2.5}) when the generator
of the cocycle is 
\begin{equation}
g_{\alpha }\left( x\right) =\exp \left( \exp \left( i\alpha \ln _{+}\left(
Df\left( x\right) \right) \right) \right)  \label{2.10}
\end{equation}
\end{definition}

As before, existence of $C\left( \alpha \right) $ depends on integrability
of $\ln _{+}\left\Vert g_{\alpha }\left( x\right) \right\Vert $ and, because 
$\exp \left( i\alpha \ln _{+}\left( Df\left( x\right) \right) \right) $ is
bounded, this is always fulfilled.

Although $C\left( \alpha \right) $ contains complete information on the
statistical properties of the local fluctuation rate a full ergodic
characterization of the dynamics should also contain information about
correlations at different points. The ergodic parameters obtained from the
Hessian in a variational formulation \cite{Carreira} already contain partial
information on the correlations, but a full study of this problem is far
from complete.

\section{Dynamical R\'{e}nyi entropies and fluctuations of the local
expansion rate}

Another way that has been used \cite{Fujisaka} \cite{Benzi} \cite%
{Grassberger1} \cite{Eckmann1} to go beyond the Lyapunov characterization is
the construction of generalized entropies.

Let $\Phi $ be a partition of $M$ and $\left\{ \phi _{i}^{(n)}\right\} $ the
elements of the partition $\Phi _{n}$ (partition refined by the dynamics $f$%
) 
\begin{equation}
\Phi _{n}=\underset{i=0}{\overset{n-1}{\vee }}f^{-i}\left( \Phi \right)
\label{3.1}
\end{equation}%
Then, the dynamical R\'{e}nyi entropy of order $\alpha $ is 
\begin{equation}
K\left( \alpha \right) =\sup_{\Phi }\left\{ \lim_{n\rightarrow \infty }\frac{%
1}{1-\alpha }\frac{1}{n}\ln \sum_{i}\mu \left( \phi _{i}^{(n)}\right)
^{\alpha }\right\}  \label{3.2}
\end{equation}%
The $\sup $ over all possible partitions (or the existence of a generating
partition) is not, in general, easy to establish. Therefore, an easier to
compute (but not necessarily equivalent) definition uses a partition of the
phase-space in uniform boxes of side $\varepsilon $ \cite{Grassberger2} \cite%
{Eckmann1}. Let the invariant measure be absolutely continuous with respect
to Lebesgue. Then, denoting by $p\left( i_{0}\cdots i_{n-1}\right) $ the
joint probability to be at the box $i_{0}$ at time $0$, to be at box $i_{1}$
at time $1$, $\cdots $, and to be at box $i_{n-1}$ at time $n-1$%
\begin{equation}
K_{B}\left( \alpha \right) =\lim_{\varepsilon \rightarrow
0}\lim_{n\rightarrow \infty }\frac{1}{1-\alpha }\frac{1}{n}\ln
\sum_{i_{0}\cdots i_{n-1}}\left( p\left( i_{0}\cdots i_{n-1}\right) \right)
^{\alpha }  \label{3.3}
\end{equation}%
the sum being over all different blocks of length $n$.

This is the definition that will be used here to obtain an estimate of its
relation to the fluctuations of the local expansion rate. The local
expansion rate $\Lambda \left( x\right) =\prod_{\lambda _{i}>0}e^{\lambda
_{i}\left( x\right) }$ of the dynamics (defined as in Section 2) implies
that if the system is in box $i_{0}$ at time $0$ it can go to $\Lambda
\left( i_{0}\right) $ boxes in the next step, then to $\Lambda \left(
i_{0}\right) \Lambda \left( i_{1}\right) $ boxes, etc. Here $\Lambda \left(
i_{k}\right) $ denotes the average expansion rate in the (small) box $i_{k}$%
. Then, one obtains for the probability $p\left( i_{0}\cdots i_{n-1}\right) $
the following estimate \cite{Eckmann1} 
\begin{equation}
p\left( i_{0}\cdots i_{n-1}\right) =\frac{\mu \left( i_{0}\right) }{\Lambda
\left( i_{o}\right) \cdots \Lambda \left( i_{n-2}\right) }  \label{3.4}
\end{equation}%
$\mu \left( i_{0}\right) $ being the measure of the $i_{0}$ box. Hence 
\begin{equation*}
K_{B}\left( \alpha \right) =\lim_{\varepsilon \rightarrow
0}\lim_{n\rightarrow \infty }\frac{1}{1-\alpha }\frac{1}{n}\ln \left(
q_{n}\left\langle \left( \frac{\mu \left( i_{0}\right) }{\Lambda \left(
i_{o}\right) \cdots \Lambda \left( i_{n-2}\right) }\right) ^{\alpha
}\right\rangle \right)
\end{equation*}%
$q_{n}$ being the number of different blocks of length $n$ and $\left\langle
\cdots \right\rangle $ denoting expectation values over blocks with this
length. $q_{n}$ is obtained from normalization $\sum_{i_{0}\cdots
i_{n-1}}p\left( i_{0}\cdots i_{n-1}\right) =1$. Then 
\begin{equation*}
K_{B}\left( \alpha \right) =\lim_{\varepsilon \rightarrow
0}\lim_{n\rightarrow \infty }\frac{1}{1-\alpha }\frac{1}{n}\ln \left(
\left\langle \frac{\mu \left( i_{0}\right) }{\Lambda \left( i_{o}\right)
\cdots \Lambda \left( i_{n-2}\right) }\right\rangle ^{-1}\left\langle \left( 
\frac{\mu \left( i_{0}\right) }{\Lambda \left( i_{o}\right) \cdots \Lambda
\left( i_{n-2}\right) }\right) ^{\alpha }\right\rangle \right)
\end{equation*}%
In the $\lim_{n\rightarrow \infty }$ one may write 
\begin{equation}
K_{B}\left( \alpha \right) =\lim_{\varepsilon \rightarrow
0}\lim_{n\rightarrow \infty }\frac{1}{1-\alpha }\frac{1}{n}\ln \left\langle
\exp \left( \left( 1-\alpha \right) \sum_{k=0}^{n-2}\ln \Lambda \left(
i_{k}\right) \right) \right\rangle  \label{3.5}
\end{equation}%
This establishes the relation between the dynamical R\'{e}nyi entropy and
what some authors \cite{Fujisaka} \cite{Eckmann1} \cite{Benzi} call \textit{%
generalized Lyapunov exponents}.

One recognizes in the above expression $\left( 1-\alpha \right) K_{B}\left(
\alpha \right) $ as the pressure function for the random variable $Y_{n}=%
\frac{1}{n}\sum_{k=0}^{n-2}\ln \Lambda \left( i_{k}\right) $ \cite{Ellis}.
Therefore, if it is differentiable (in $\alpha $), its Legendre transform 
\begin{equation}
I\left( y\right) =\sup_{\alpha }\left\{ \left( 1-\alpha \right) y-\left(
1-\alpha \right) K_{B}\left( \alpha \right) \right\}  \label{3.6}
\end{equation}
is the deviation function for the large deviations of the random variable $%
Y_{n}=\frac{1}{n}\sum_{k=0}^{n-2}\ln \Lambda \left( i_{k}\right) $, that is,
it characterizes the probability $P_{n}$ for finite-time fluctuations in the
computation of the sum of the positive Lyapunov exponents. 
\begin{equation*}
P_{n}\left\{ \frac{1}{n}\sum_{k=0}^{n-2}\ln \Lambda \left( i_{k}\right) \in
\left( y,y+dy\right) \right\} \asymp \exp \left( -nI\left( y\right) \right)
dy
\end{equation*}
the symbol $\asymp $ meaning logarithmic equivalence.

This establishes a general relation between the dynamical R\'{e}nyi entropy
and the fluctuations of the local expansion rate. Under more strict
conditions, that is, if the correlation between successive values of $%
\Lambda \left( i_{k}\right) $ decays sufficiently fast, namely if 
\begin{equation}
\left\langle \exp \left( \left( 1-\alpha \right) \sum_{k=0}^{n-2}\ln \Lambda
\left( i_{k}\right) \right) \right\rangle \prod_{k=0}^{n-2}\left\langle \exp
\left( \left( 1-\alpha \right) \ln \Lambda \left( i_{k}\right) \right)
\right\rangle ^{-1}\leq c_{1}e^{c_{2}n^{\gamma }}  \label{3.7}
\end{equation}%
with $c_{2}>0$ and $\gamma <1$, then 
\begin{equation}
K_{B}\left( \alpha \right) =\lim_{\varepsilon \rightarrow 0}\frac{1}{%
1-\alpha }\ln \left\langle \exp \left( \left( 1-\alpha \right) \ln \Lambda
\left( i\right) \right) \right\rangle  \label{3.8}
\end{equation}%
which one recognizes as a cumulant generating function. Summarizing:

\begin{proposition}
The Legendre transform of the (box) dynamical R\'{e}nyi entropy is the
deviation function of the local expansion rate. If the weak correlation
condition (\ref{3.7}) is verified then 
\begin{equation}
K_{B}\left( \alpha \right) =\lim_{\varepsilon \rightarrow
0}\sum_{s=1}^{\infty }k_{s}\left( \ln \Lambda \right) \left( 1-\alpha
\right) ^{s-1}  \label{3.9}
\end{equation}
where $k_{s}\left( \ln \Lambda \right) $ are the cumulants of the local
expansion rate.
\end{proposition}

In its range of validity Eq.(\ref{3.9}) is a generalization of Pesin's
formula \cite{Pesin1}. Grassberger and Procaccia \cite{Grassberger3} have
also obtained a similar, although more complex, relation between the
dynamical R\'{e}nyi entropy and the fluctuations of the expansion rate.

\section{Ergodic parameters and measures of complexity}

To have quantitative measures of complexity and self-organization is an
important issue for a mathematical theory of complex systems. The \textit{%
algorithmic complexity} \cite{Kolmogorov} \cite{Chaitin} of the signal
generated by a dynamical system, that is, of the sequence of numbers coding
a particular orbit, is the limit 
\begin{equation}
C_{K}(S)=\lim_{n\rightarrow \infty }\frac{M_{n}(S)}{n}  \label{5.1}
\end{equation}
where $M_{n}(S)$ is the length of the smallest program (code plus data) able
to generate the first $n$ symbols of the sequence. Up to a factor, the
average algorithmic complexity of the sequences is identical to the \textit{%
Shannon entropy }\cite{Shannon} of the system considered as a source
emitting the sequence.

The notion of algorithmic complexity applies to each particular sequence,
whereas an ergodic invariant like the \textit{Kolmogorov-Sinai entropy} is a
statistical parameter referring to the average behavior of the orbits.
Nevertheless, the two notions are related. Let in $M_{n}(S)$ distinguish two
components 
\begin{equation}
M_{n}(S)=c_{1}(n)+c_{2}n  \label{5.2}
\end{equation}
where $c_{1}(n)$ is the length of the code and $c_{2}n$ the size of the
input data. $c_{2}n$ is the part of the information that is not explained by
the program code. Therefore, as far as the model program is concerned, $%
c_{2}n$ is the random component of the sequence. In general $\frac{c_{1}(n)}{%
n}\rightarrow 0$ when $n\rightarrow \infty $ and only the random component
contributes to the algorithmic complexity. For this reason, in many cases,
the algorithmic complexity of typical orbits coincides (up to a factor) with
the Kolmogorov-Sinai entropy \cite{Brudno} \cite{White}.

The algorithmic complexity, the Shannon entropy and the Kolmogorov-Sinai
entropy (rate) measure the degree of unpredictability (or irregularity) of
the system but not necessarily the difficulty of modelling it from
experimental observations. In fact a system generating completely random
sequences has maximum algorithm complexity, but may be modelled by a simple
random number generator.

A better characterization of what is usually meant by complexity is the
notion of \textit{excess entropy} \cite{Crutchfield0} or \textit{effective
measure complexity} \cite{Grassberger1} \cite{Grassberger2}. Let $%
p_{N}(s_{1}\cdots s_{n})$ be the probability to find the block $s_{1}\cdots
s_{n}$ of size $n$. Then 
\begin{equation}
H(n)=-\sum_{\{s_{i}\}}p_{n}(s_{1}\cdots s_{n})\log p_{n}(s_{1}\cdots s_{n})
\label{5.3}
\end{equation}
and 
\begin{equation}
h_{s}=\lim_{n\rightarrow \infty }\frac{1}{n}H(n)  \label{5.4}
\end{equation}
is the \textit{Shannon entropy}.

The difference $\frac{1}{n}H(n)-h_{s}$ represents the additional information
(beyond the one obtained from size $n$ blocks) that is needed to reveal the
true long-term unpredictability of the system. Summing all these
differences, the \textit{excess entropy }$E$ grows with the amount of effort
(and time) that is needed to construct an accurate model of the system. 
\begin{equation}
E=\sum_{n}\left( \frac{1}{n}H\left( n\right) -h_{s}\right)  \label{5.4a}
\end{equation}
It is a measure of the diversity of dynamical structures that is present in
the information source. The nature of the information processing employed by
the dynamical system to produce its unpredictability is captured by the 
\textit{statistical complexity} $C_{s}$ \cite{Crutchfield1} \cite%
{Crutchfield2}, related to the excess entropy by 
\begin{equation}
E\leq C_{s}  \label{5.5}
\end{equation}
meaning that, given an event, the ideal prediction of another one requires
an amount of information at least equal to the mutual information between
the two events.

The Kolmogorov-Sinai entropy, bounded by the sum of the positive Lyapunov
exponents (an infinite-time average), is a measure of the complexity of
typical orbits. On the other hand , it is to be expected that the
finite-time fluctuations in the calculation of the Lyapunov exponents be a
symptom of the diversity of dynamical structures. Therefore these
fluctuations might be related to the excess entropy and therefore be a
measure of the dynamical complexity of the system. Here such a relation is
established.

One uses the large deviation principle that states that the Legendre
transform $I\left( y\right) $ (Eq.(\ref{3.6})) of $\left( 1-\alpha \right)
K_{B}\left( \alpha \right) $ (in Eq.(\ref{3.5})) is the deviation function
of the random variable $Y_{n}=\frac{1}{n}\sum_{k=0}^{n-2}\ln \Lambda \left(
i_{k}\right) $. For invariant measures absolutely continuous with respect to
Lebesgue, the average value of $Y_{n}$ is an estimate of $\frac{1}{n}H\left(
n\right) $. Therefore one may write 
\begin{equation}
E_{e}=\sum_{n}\left\{ \int_{0}^{\infty }yP_{n}\left( y\right) dy-y_{I_{\min
}}\right\}  \label{5.6}
\end{equation}
with $y_{I_{\min }}$ being the value that minimizes $I\left( y\right) $ and 
\begin{equation}
P_{n}\left( y\right) =\frac{e^{-nI\left( y\right) }}{\int_{0}^{\infty
}e^{-nI\left( y\right) }dy}  \label{5.7}
\end{equation}

One sees that a \textit{dynamical complexity measure} $E_{e}$ analogous to
the excess entropy $E$ may be computed from the ergodic parameters that
define the fluctuations of the local expansion rate.

In Ref.\cite{Shalizi}, the authors have proposed to measure the increase of 
\textit{self-organization} between time $t_{1}$ and $t_{2}$ by the change in
the statistical complexity 
\begin{equation}
\Delta C_{s}=C_{s}\left( t_{2}\right) -C_{s}\left( t_{1}\right)  \label{5.8}
\end{equation}
if this is not due to the action of an external agent. Given the relation (%
\ref{5.5}) this might also be estimated by the change of excess entropy.

This might be an appropriate notion when one is comparing two different
states of an evolving system. There is however another aspect of what is
usually understood as self-organization in multi-agent systems that relates
to the interrelation between the dynamics of the agents (and their local
cluster) and the global collective dynamics. This aspect is better
characterized by the relation between the Lyapunov exponents and the
conditional ones (see \cite{Vilela1} and \cite{Vilela4} for details).

\end{document}